\documentclass{article}

 \usepackage[creativeai, final]{neurips_2025}

\usepackage[utf8]{inputenc} 
\usepackage[T1]{fontenc}    
\usepackage{hyperref}       
\usepackage{url}            
\usepackage{booktabs}       
\usepackage{amsfonts}       
\usepackage{nicefrac}       
\usepackage{microtype}      
\usepackage{xcolor}         
\usepackage{graphicx}
\usepackage{tabularx}
\usepackage{amsmath}
\usepackage{amssymb}
\title{Go witheFlow: Real-time Emotion Driven Audio Effects Modulation}

\author{
  Edmund Dervakos\thanks{Equal contribution.} \\
  \texttt{eddiedervakos@islab.ntua.gr}
  \And
  Spyridon Kantarelis\footnotemark[1] \\
  \texttt{spyroskanta@ails.ece.ntua.gr}
  \And
  Vassilis Lyberatos\footnotemark[1] \\
  \texttt{vaslyb@ails.ece.ntua.gr}
  \And
  Jason Liartis \\
  \texttt{jliartis@ails.ece.ntua.gr}
  \And
  Giorgos Stamou \\
  \texttt{gstam@cs.ntua.gr} \\
  \\
  Artificial Intelligence and Learning Systems Laboratory\\
  National Technical University of Athens, Athens, Greece
}

\begin{document}

\maketitle

\begin{abstract}
  Music performance is a distinctly human activity, intrinsically linked to the performer’s ability to convey, evoke, or express emotion. Machines cannot perform music in the human sense, they can produce, reproduce, execute or synthesize music, but they lack the capacity for affective or emotional experience. As such, music performance is an ideal candidate through which to explore aspects of collaboration between humans and machines. In this paper, we introduce the witheFlow system, designed to enhance real-time music performance by automatically modulating audio effects based on features extracted from both biosignals and the audio itself. The system, currently in a proof-of-concept phase, is designed to be lightweight, able to run locally on a laptop, and is open-source given the availability of a compatible Digital Audio Workstation, and sensors.
\end{abstract}

\section{Introduction}

As AI is increasingly dominating creative domains, often positioned as an autonomous creator or composer \cite{civit2022systematic}, we should investigate alternative forms of AI-human collaboration that preserve the human role in artistic expression, instead of treating AI as a replacement for human creativity \cite{mahmud2023study}. AI offers the ability to serve as a sophisticated tool, handling technical details and complex processing tasks, giving humans the space to remain at the center of creation, generating aesthetic ideas and making artistic decisions that define the creative vision~\cite{mccormack2019autonomy}. This collaboration creates a more liberated environment where performers can focus on self-expression, using AI as an enhancing tool rather than a creative competitor~\cite{carter2017using}.

\begin{figure}[!ht]
    \centering
    \includegraphics[width=0.9\linewidth]{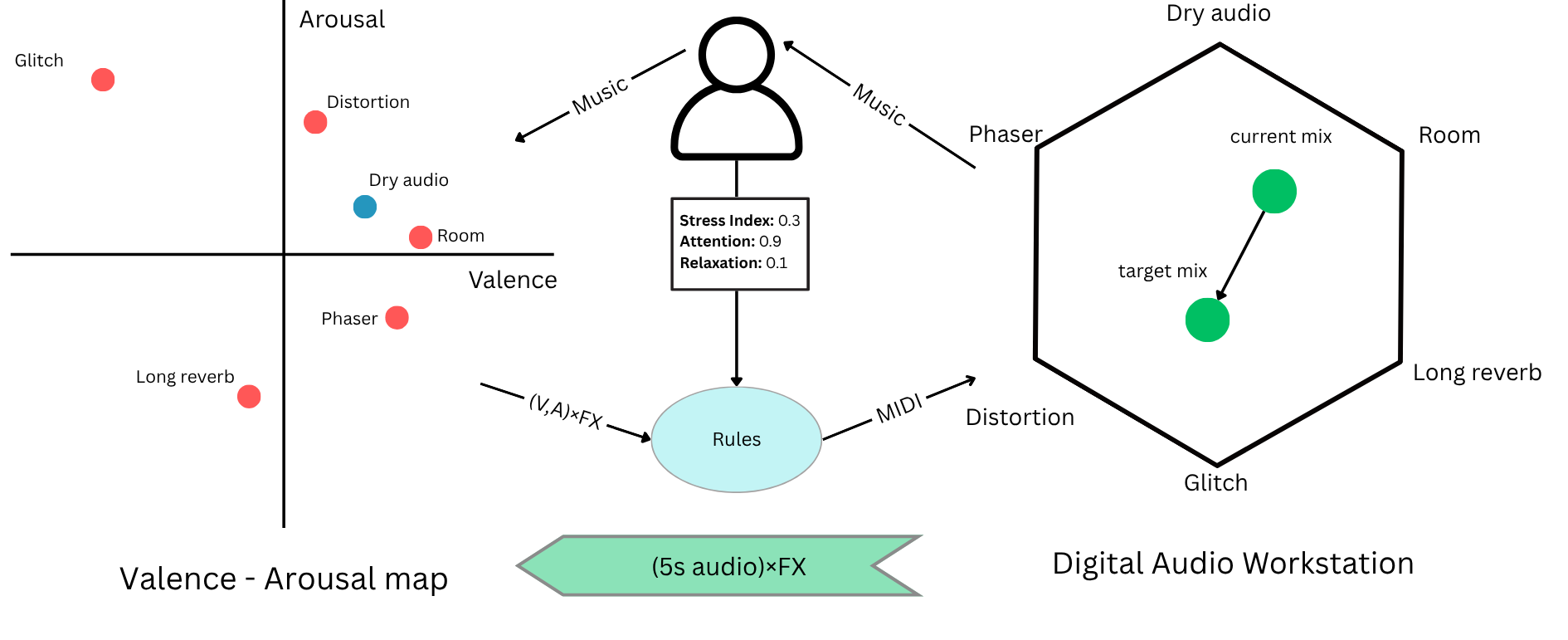}
    \caption{Overview of how audio Valence-Arousal (VA) interact with audio effects based on biosignals.}
    \label{fig:overview}
\end{figure}


Creative AI in music has predominantly focused on music generation~\cite{zhu2023survey}, while traditional AI applications have concentrated on classification~\cite{ndou2021music}, analysis~\cite{yust2022clustering} and recommendation tasks~\cite{zhang2023symbolic_classification}. The concept of using AI as a live-performance or studying assistant remains underexplored in the current landscape \cite{10.1145/3706599.3720052}. Our main idea centers on utilizing lightweight models that perform locally in real-time, offering instant feedback to live performers or music students without the latency and computational overhead. Our development of such a system as a proof of concept demonstrates how AI can leverage these strengths alongside existing Digital Audio Workstation (DAW) systems to enhance rather than replace human musical creativity.


Traditional music technology primarily processes audio signals but overlooks the performer's real-time emotional and physiological state. AI systems can bridge this gap by processing biosensor data to understand performers' emotional states and translate them into responsive audio modifications~\cite{lyberatos2025musicinterpretationemotionperception}. By integrating this physiological feedback, we create a more complete picture of the performer's expressive intent, enabling adaptive audio processing that responds to individual performers rather than imposing standardized responses~\cite{gruzelier2014eeg}. This approach opens possibilities for new forms of musical expression where performers can develop enhanced bio-awareness and mind-body integration, creating a direct connection between their internal emotional experience and the sonic output of their performance.

We present the \textbf{witheFlow} system, it combines lightweight machine learning models with traditional rule-based AI to process multiple input streams from performers in real-time. An overview of our approach is depicted in Figure~\ref{fig:overview}. It integrates biosensor data and audio analysis, allowing the system to respond to both physiological and musical cues simultaneously. Rather than altering musical content, our primary focus is on dynamically adjusting audio effects and processing parameters \cite{tsiros2020towards} as tools for enhancing expressivity and achieving alignment between the performer's emotional state and their sonic output. A demo video of the witheflow system can be found on youtube~\footnote{\url{https://www.youtube.com/watch?v=oBbYrvduU2k}}.

\section{The witheFlow System}
\begin{figure}[!ht]
    \centering
    \includegraphics[width=0.9\linewidth]{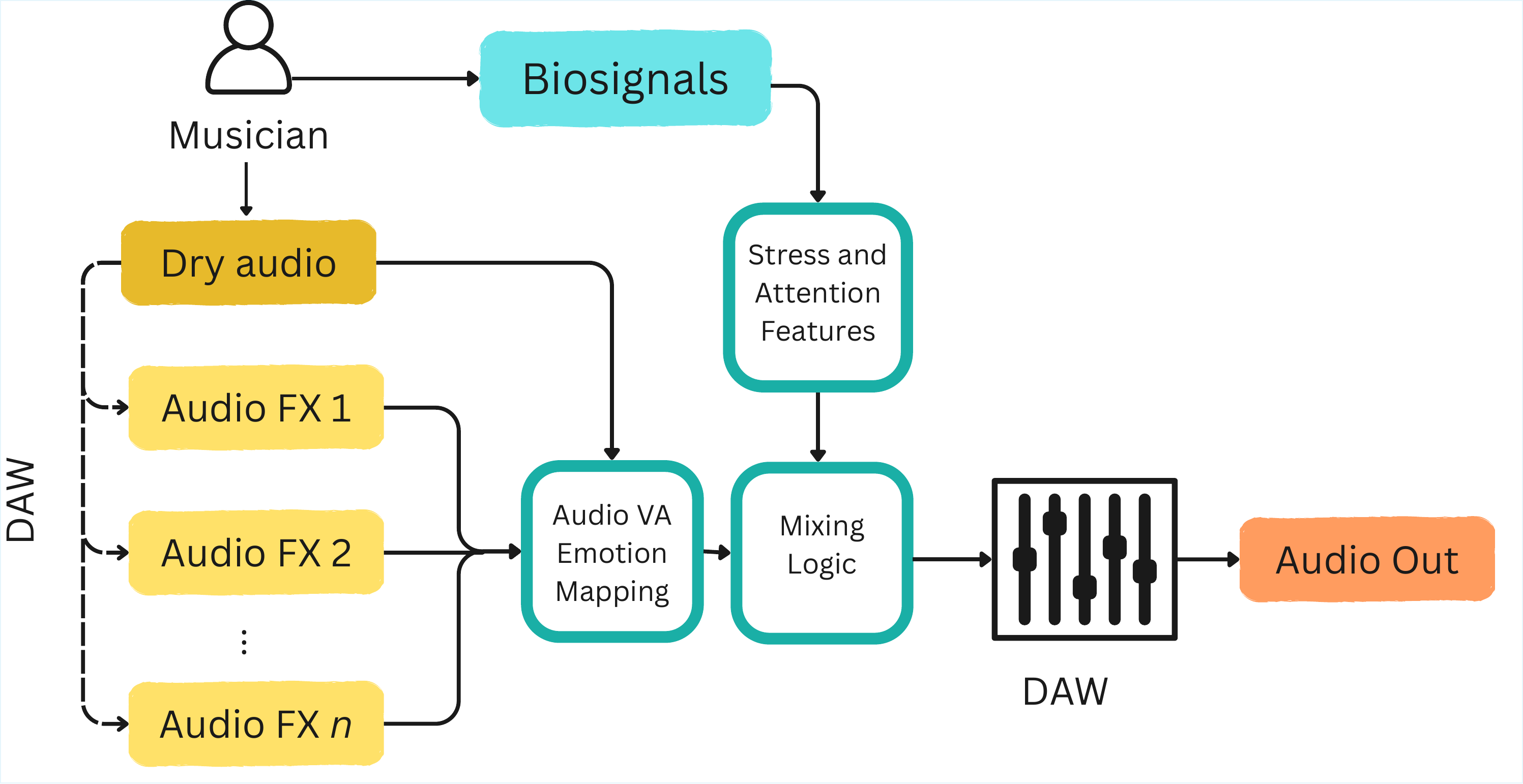}
    \caption{The witheFlow System.}
    \label{fig:archi}
\end{figure}

The overall architecture of the system is depicted in Figure \ref{fig:archi}. The three main components are: a) a biosignal-based `emotional state' feature extractor, b) an audio-based emotion regressor in the Valence-Arousal (VA) space, and c) a rule-based mixing logic module. The system is implemented in Python, with inter-module communication facilitated via MIDI (Musical Instrument Digital Interface) protocol messages \cite{rothstein1995midi}. 

\paragraph{Preparation}
The performer creates up to \textit{n} effects chains that are each routed to separate channels in the DAW. These can be anything, from their own pedalboard, to a complicated plugin chain. The number of channels depends on the hardware. Also, calibration, and a sanity check are performed on biosensors before enabling the appropriate rule-set.

\paragraph{Biosignals}

We employ commercial-grade electroencephalography (EEG) and electrocardiography (ECG) sensors to estimate emotional state as a tuple of two physiological dimensions: \textbf{Attention/Relaxation}, derived from EEG, and the \textbf{Baevsky Stress Index (SI)}, computed from ECG. The Baevsky SI is a well-established heart rate variability (HRV) metric that quantifies stress by reflecting the balance between sympathetic and parasympathetic nervous system activity \cite{sahoo2019stress}.

The EEG sensor uses four electrodes placed at standard sites: O1, O2, T3, and T4. Signals are sampled at 250\,Hz, and power is extracted in the Alpha (8--13\,Hz) and Beta (13--30\,Hz) bands. Attention and relaxation are computed over sliding windows of 1000 samples (4 seconds). Increased beta power correlates with heightened attention~\cite{palacios2021increase}, while elevated alpha power indicates greater relaxation~\cite{sugimoto2024electroencephalographic}. These metrics are defined as:

\[
\text{Attention} = \frac{\text{Beta Power}}{\text{Alpha Power} + \text{Beta Power}}, \quad
\text{Relaxation} = \frac{\text{Alpha Power}}{\text{Alpha Power} + \text{Beta Power}}.
\]

The ECG signal is sampled at 1000\,Hz. Stress is estimated using a sliding window approach: every 0.5 seconds (500 samples), RR intervals from the preceding 15 seconds (30 windows) are analyzed to compute the Baevsky SI. This index enables continuous assessment of autonomic nervous system balance and stress reactivity. The Baevsky SI is defined as:

\[
SI = \frac{AMo \times 100\%}{2 \times Mo \times MxDMn},
\]

where \(AMo\) is the amplitude of the mode (percentage of RR intervals within the modal bin), \(Mo\) is the mode of RR intervals (most frequent RR interval in milliseconds), and \(MxDMn = RR_{\max} - RR_{\min}\) is the variation range of RR intervals (in milliseconds).

\paragraph{Audio Routing}
The application of audio effects and the routing of audio to different channels is done through a DAW. In particular, the performer's dry audio (with no effects applied) is routed through all effects channels in parallel and all channels (dry and effects) are routed through a virtual audio device to be available in Python via the sounddevice package\footnote{\url{https://python-sounddevice.readthedocs.io/}}. All channels are also routed to their own `monitor' channel that then gets sent to the system's audio output. The monitor channels are all muted by default, and their gain is controlled by the mixing logic.  

\paragraph{Audio Emotion Regressor} 
In the audio domain, we model emotions based on Russell's circumplex model of valence and arousal \cite{russell1980circumplex}. We start with a PANNs CNN10 \cite{kong2020panns} that we trained on the DEAM dataset \cite{aljanaki2017developing}, after deleting the final classification layer and replacing it with a linear regressor with two outputs (arousal and valence). The audio model has a window size corresponding to 5 seconds, while the audio itself is down-sampled to 30kHz. We have not done any significant hyperparameter tuning, but are in the process of developing a multi-instrumental dataset of solo performances that will facilitate this part of the process, as datasets typically contain full music productions and not solo performances (see Section \ref{sec:discussion}).

\paragraph{Mixing Logic}
The mixing logic determines the contribution of each channel, including the dry audio, to the final mix by adjusting their gains.
It operates through a set of rules that are applied dynamically based on a combination of the musician’s stress and attention levels, and the valence-arousal characteristics of the dry audio.
Channel gains are modified according to their VA values to shape the overall valence and arousal of the output mix.

In our implementation, the mixing logic is fully customizable.
It comprises rulesets encoded in YAML \footnote{\url{https://yaml.org}} files, each consisting of a set of conditions of the form $a<x<b$, where $x$ is one of stress, attention, valence, or arousal (of the dry signal). Each condition is paired with a function, triggered when the firing conditions are satisfied, and a textual description.
We currently have defined four rulesets, three for different combinations of the available biosignal sensors, and one for just the audio.
The functions are implemented in Python, and each boosts the gain of a different effect. For example, one of the functions boosts the gain of the closest effects channel in VA space, one boosts the furthest, another boosts the lowest arousal channel, etc.
The rules can easily be customized by the end user by providing their own YAML files and/or defining their own Python functions. The ruleset for the case of all sensors being available is shown in table \ref{tab:va_mixing_logic}. For a detailed description of each ruleset and the functions, we refer to our GitHub repository \footnote{\url{https://github.com/witheflow/system}}.
In the future we aim to explore approaches for a learnable mixing logic.


\begin{table}[!ht]
\centering
\caption{Mixing logic based on biosignal quadrant and affective distance in Valence-Arousal (VA) space.}
\begin{tabular}{ccp{10cm}}
\toprule
\textbf{Stress} & \textbf{Attention} & \textbf{Gain Logic (in VA space)} \\
\midrule
High & High & Boost FX that are far from the dry signal and lie in the direction of increasing arousal (i.e., arousal\textsubscript{fx} > arousal\textsubscript{dry}). Suppress FX in other directions. \\
High & Low & Boost FX that are far from the dry signal, regardless of direction. Suppress FX that are close to the dry signal. \\
Low & High & Boost FX that are close to the dry signal in VA space. Suppress FX that are far. \\
Low & Low & Boost FX that are close to the dry signal and have lower arousal than the dry signal (i.e., arousal\textsubscript{fx} < arousal\textsubscript{dry}). Suppress FX that are far or increase arousal. \\
\bottomrule
\end{tabular}
\label{tab:va_mixing_logic}
\end{table}

The intuition behind the rules is the following.
If high stress is detected, then we boost the gain of FX that are furthest in the VA space than the current state, with the assumption that the musician's stress is somehow expressed via their performance, so we aim for audio that is as different as possible from the current audio.
If stress is low, we stay close to the current audio.
Regarding attention, in our experimentation, it can be controlled, to an extent, by the performer, so in that sense we interpret it as an indicator of intent.
Thus, if the current state is high arousal, we aim for higher arousal audio in our mix, whereas if it is relaxed, we aim for lower arousal audio. 

We can formally describe the mixing logic as a piecewise function over an input domain $\mathcal{D}$ --- in our case, the stress, attention, and VA space values --- the partition of the input domain $\mathcal{P} = \{\mathcal{D}_1, \mathcal{D}_2, \dots, \mathcal{D}_m\}$ s.t. $\bigcup_{i = 1}^m \mathcal{D}_i = \mathcal{D},\,  \mathcal{D}_i \cap \mathcal{D}_j = \varnothing,\, \forall i, j = 1, \dots, m$, an output codomain $\mathcal{G}$ --- in our case, the gains of each FX channel, which is a vector of $n$ positive real values --- and a set of sub-functions $f_1, f_2, \dots, f_m$, with $f_i: \mathcal{D}_i \rightarrow \mathcal{G}$, which define the gain logic.

Although the partition of the input domain could be arbitrary, partitions that can be described as a list of rules or a decision tree are preferred because they are simple and transparent.
The sub-functions could also be arbitrary, but in our implementation we have a set of predefined functions that can be assigned to each subdomain.
This formalization is intentionally chosen to align with our future goal of learning the mixing logic from data. By structuring the logic as a partitioned input space with interpretable rule-based mappings, we pave the way for training decision trees or similar symbolic models that learn the subdomains, and the sub-function to apply to each subdomain, so that they mimic or refine user-defined rules while retaining interpretability and modularity.


\paragraph{Robustness}

Artifact detection is applied to both EEG and ECG signals. Prolonged low signal values are considered invalid. When a persistent artifact is detected—typically indicating poor device contact—the affected device is deactivated. For transient EEG artifacts, features are extracted from the remaining functional electrodes. If all electrodes fail, feature values are imputed from previous time steps. Additionally, when an artifact is detected, the system dynamically updates its rule set. This enables adaptive mixing logic that prioritizes the most reliable signal sources available at any given time. Finally, the strength of the rules can be controlled by the performer via MIDI (for example with a foot pedal), and can even be reversed, giving the performer additional controllability.

\section{Discussion and Research Directions}
\label{sec:discussion}

The proposed system has been co-developed and tested in collaboration with multiple musicians, who have reported a generally positive experience, particularly during improvisational sessions. This collaborative process has informed both the design and refinement of the system. Several research directions remain open, aimed at advancing the system toward a more capable and responsive AI collaborator for musical performance, including the development of a formal framework for quantitative evaluation.

First of all, datasets are crucial for any system that utilizes machine learning. Music is a particularly challenging domain, due to copyrights, and the sensitive nature of the data itself. Even though numerous datasets exist that explore music and emotions, there are few that focus on solo performance and include real-time annotations. The addition of biosignals, which are considered personal health information, makes the development of such datasets even more challenging. Part of our ongoing research while developing witheFlow involves collaborating with professional and amateur musicians with the goal of creating and publishing such a dataset (preliminary analysis is published here~\cite{lyberatos2025musicinterpretationemotionperception}).   

Secondly, the representations that we choose for the data, the features that we extract, and the training objectives that we set are equally important for the end result. In the proof-of-concept presented in this paper we represent physiological state as Attention/Relaxation and Stress, and audio as Valence and Arousal. Furthermore, the sensors that we used throughout the development have specific constraints and specifications regarding available data and its quality. Thus, an interesting research direction involves an interdisciplinary exploration of extractable features that are meaningful and reliable in the context of emotional expression during music performance, not only from audio signals and biosignals, but from more types of sensors and modalities, such as video. Such features can then be used for training machine learning models on downstream tasks that can be utilized in a system such as witheFlow.

Having meaningful and understandable features is crucial if the machine learning components are to be interpretable and controllable, which we believe are important attributes for such a system.  In particular, explainability and controllability are essential not only from a technical or usability perspective, but also as a mode of communication between the performer and the system. Making the decision-making processes of the system transparent (e.g., which rule was triggered, or which physiological signal caused a change in mixing) helps establish trust, as the system is communicating its behavior to the user~\cite{lyberatos2025challenges}. Furthermore, allowing the user to control or steer the system serves this communication in the other direction. As such, one direction to follow is developing an interpretable-by-design, and controllable machine learning pipeline as a mixing logic for witheFlow, instead of having hard-coded rules.

At the same time, black-box end-to-end architectures that operate directly on raw biosignals and audio - for example, deep reinforcement learning agents trained to modulate effects for minimizing stress, or utilizing approaches such as DDSP~\cite{engel2020ddsp} and MIDI-DDSP \cite{wu2021midi} for incorporating the audio manipulation in the end-to-end architecture- represent an equally compelling direction. These could offer novel forms of expressivity, potentially discovering relationships and behaviors that would be difficult to otherwise engineer.

Finally, an important consideration lies in the hardware and software architecture that underpins such a system. Real-time interaction imposes strict constraints on latency, reliability, and usability, especially in performance contexts. A local setup, such as the one used in our current implementation, offers several advantages: it supports low-latency interaction, avoids the need for internet connectivity, respects the privacy of biosignal data, and allows musicians to use the system in a self-contained, portable form. It also aligns well with energy-efficient, edge-compatible models, making it accessible in diverse performance settings.
However, as models grow in size and complexity—particularly with the emergence of foundation models for music or large-scale transformers that incorporate multimodal context—cloud-based computation becomes a viable and powerful option. These architectures allow access to cutting-edge models that may be too large or resource-intensive to run locally, enabling more sophisticated inference and potentially richer creative behavior. At the same time, they raise challenges around data privacy, network latency, and system robustness, especially in live performance scenarios where timing is critical.
A promising direction might involve hybrid or adaptive architectures, where smaller interpretable models run locally and handle real-time interaction, while larger cloud-based systems provide higher-level insights, suggestions, or periodic adjustments. Balancing these trade-offs will be essential in the future evolution of witheFlow and similar emotionally adaptive creative systems.  

\section{Ethical considerations}

This study was conducted in accordance with the ethical principles outlined in the Declaration of Helsinki. All participants provided informed consent prior to their participation, and their anonymity was maintained throughout the study. Ethical approval was obtained from the local ethics committee of the National and Technical University of Athens. Participants were informed of their right to withdraw at any time, and all data were anonymized and securely stored to ensure confidentiality.

The development of bio-responsive musical systems raises important ethical considerations regarding sensitive personal data. Biosignals and original music performances require careful handling and robust privacy protections. While our current proof of concept operates locally and minimizes data exposure, scaling would likely require cloud computation and data collection, introducing risks of data mismanagement and biased model development. Current audio classification models are trained on narrow datasets that may not represent diverse musical expressions, and valence-arousal annotations are inherently subjective. Despite the system's role in shaping sonic output, authorship of the musical performance belongs entirely to the performer. The musicians who participate in the development of witheFlow are recruited individually and compensated fairly for their performance time. They are asked to wear non-invasive biosensors while performing. No written instructions are provided; all directions are communicated verbally in person. Care is taken to ensure that the sensors do not cause discomfort and do not interfere with their ability to perform.

Finally, our system can be interpreted as a closed loop: the musician’s internal state influences the audio FX applied in real time, and those changes in sound can, in turn, shape their psychological response. While this dynamic can aid creative expression, it should also be carefully considered in terms of psychological safety. To avoid unwanted effects or loss of control, the artist can personally configure the mixing logic and an immediate override (via foot pedal) enables them to reverse the effects. We believe that systems like this should prioritize the artist’s ability to intervene or opt out, especially when working with sensitive physiological inputs.

\section{Conclusion}
This work introduces the witheFlow system, it integrates biosignals with audio processing to enhance musical performance while maintaining human creative agency. It offers an implemented prototype, technical design, and conceptual framing for integrating biosignals and emotion-aware audio processing in live performance contexts. In our experience, musicians generally enjoy interacting with the system, particularly in improvisational settings, where the system’s responsiveness to their internal and musical state opens up new expressive possibilities. As we continue developing an evaluation framework tailored to such systems, we believe that a robust and flexible methodology will help researchers and artists explore, refine, and compare diverse approaches to co-creative human-AI interaction—contributing to the broader evolution of emotionally adaptive music technologies.

\section*{Acknowledgements}

The research project is implemented in the framework of H.F.R.I call “Basic research Financing (Horizontal support of all Sciences)” under the National Recovery and Resilience Plan “Greece 2.0” funded by the European Union –NextGenerationEU (H.F.R.I. Project Number: 15111 - Emotional Artificial Intelligence in Music Expression).

\bibliographystyle{plain}  
\bibliography{references}

\end{document}